\documentclass[traditabstract]{aa} 
\usepackage{graphicx}
\usepackage{txfonts}
\usepackage{subfig}

\begin{document}

   \title{An investigation of magnetic field distortions
          \\ in accretion discs around neutron stars}

   \subtitle{II. Analysis of the toroidal field component}

   \author{L.~Naso\inst{\ref{inst1}}\and
          J.~C.~Miller\inst{\ref{inst2}}\fnmsep\inst{\ref{inst3}}
          }

   \institute{
Key Laboratory of Solar Activity, National Astronomical Observatories, Chinese
Academy of Sciences, \\
20A Datun Road, Chaoyang District, Beijing 100012, China\\
\email{luca.naso@gmail.com}\label{inst1}
         \and
SISSA and INFN, via Bonomea 265, I-34136 Trieste, Italy\label{inst2}
         \and
             Department of Physics (Astrophysics), University of Oxford, Keble
Road, Oxford OX1 3RH, UK\label{inst3}
             }

 \date{Received 14 December 2010 / Accepted 4 April 2011 }

 
  \abstract {Millisecond pulsars are believed to be old pulsars spun up by a
surrounding accretion disc. Magnetic fields are thought to play a leading role
in this, both by determining the location of the inner edge of the disc and by
exerting an additional torque on the star (as a result of the interaction
between the stellar magnetic field and the disc plasma motion, which creates a
toroidal component $B_\phi$). In some well-known analytic models, developed in
the 1980s, the $B_\phi$ profile was taken to be proportional to the relative
angular velocity between the disc plasma and the neutron star, multiplied by a
vertical dipolar field. The present work stands in the line of improving those
models, suggesting a new profile for $\mathbf{B}$. In a previous paper, we
discussed the poloidal component of the magnetic field and here we consider the
toroidal component, again making the kinematic approximation and looking for
steady solutions of the induction equation for axisymmetric models. The poloidal
magnetic field is not assumed to be dipolar and the poloidal velocity field is
not taken to be zero everywhere. We also do not use the thin disc approximation
to simplify the induction equation but instead solve it numerically in full 2D.
The profile obtained in the earlier analytic models is shown to have very
limited validity and a more general semi-analytic solution is proposed.}


 \titlerunning{Magnetic field distortions in accretion discs around neutron
stars. II.}
 \authorrunning{L.~Naso and J.~C.~Miller}

   \maketitle
%


\section{Introduction}\label{sec:INTRO}
 In the present work we study the deformation caused in a neutron star's 
magnetic field because of the interaction with the matter in a surrounding 
accretion disc. A basic description for this kind of system was given by Ghosh 
and Lamb in \cite{GL79a}, with the model subsequently being improved by Wang
(\cite{W87}) and Campbell (\cite{C87}), who suggested an analytic expression for
the toroidal component of the field. This expression has been widely used since
then, on account of it being both simple and physically plausible.

In these analytic models, the authors made the kinematic approximation, looking 
for an axisymmetric stationary solution of the induction equation with a given 
unchanging structure for the fluid in the disc. They further took the disc to 
be thin, the poloidal component of the magnetic field to be exactly dipolar, 
and the velocity field to have zero poloidal component, with its azimuthal 
component being Keplerian inside the disc\footnote{Campbell also considered 
non-Keplerian flow in the inner part of the disc.} and corotating at the top of 
the disc. Using cylindrical coordinates {($\varpi$, $\phi$, $z$)}, they found
 \begin{equation}
\label{eq:bp_an}
B_\phi = \gamma_{\rm a} \, (\Omega_K - \Omega_{\rm s}) \, B_z
\, \tau_{\rm d} \propto \Delta \Omega / \varpi^3 \, {\rm ,}
\end{equation}
 where $\gamma_{\rm a}$ is the amplification factor, $\Omega_K$ and $\Omega_{\rm
s}$ are the Keplerian and stellar angular velocities, respectively, and
$\tau_{\rm d}$ is the dissipation time scale. The amplification factor
$\gamma_{\rm a}$ was taken to be a constant not much greater than $1$ (it
depends on the steepness of the transition -- in the z-direction -- between
Keplerian motion inside the disc and corotation with the star at the top of the
disc). The precise profile of $\tau_{\rm d}$ depends on what is the dominant
mechanism for dissipating the magnetic field. Wang (\cite{W95}) considered three
different cases, with $\tau_{\rm d}$ being dominated by the Alfven velocity,
turbulent diffusion and magnetic reconnection, respectively.

Equation (\ref{eq:bp_an}) was then used for calculating the net magnetic torque
exerted on the neutron star. The vertically averaged torque can be written as
 \begin{equation}
 \label{eq:tor}
 \bar{\Gamma} \propto B_\phi \, B_z \, \varpi / h \, {\rm ,}
\end{equation} 
 where $h$ is the semi-thickness of the disc. Regions of the disc inward of the
corotation point therefore give positive contributions to the torque (because
$B_\phi>0$), while the remainder of the disc gives negative contributions
(because $B_\phi<0$). The total magnetic torque is obtained by integrating the
local values from the inner edge to the outer boundary, and it can be either
positive or negative depending on the location of the inner edge of the disc
with respect to the corotation point.

The aim of the present paper is to develop a semi-analytic model that can 
improve on those of Wang and Campbell, while remaining simple enough to be 
useful for people discussing the behaviour of astrophysical sources, giving a 
conceptual picture to go alongside results from large numerical calculations 
where the full set of the MHD equations is solved.

\begin{figure}[ht]
 \centering
 \includegraphics[width=0.4\textwidth]{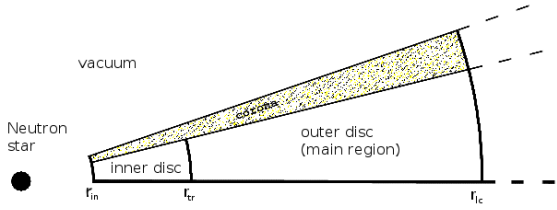}
 \caption{Schematic representation of our model (not to scale). We use $r_{\rm
in} = 10\,r_{{\rm g}}$, $r_{\rm tr} \sim 22 \, r_{{\rm g}}$ and $r_{\rm lc} \sim
115\,r_{{\rm g}}$. The opening angles are $8^{\circ}$ for the disc alone and
$10^{\circ}$ for the disc plus corona. The outer disc extends much further out
than the main region shown here: the grid continues until $r_{\rm out} = 380 \,
r_{{\rm g}}$.}
 \label{fig:model}
\end{figure}

Our approach, for the time being, is to continue to retain axisymmetry and the 
kinematic approximation but to calculate a consistent steady-state solution for
the magnetic field, relaxing the assumptions on the poloidal components of the
magnetic and velocity fields and using a 2D model without any vertical averaging
of the Taylor expansion of the induction equation. In the main region of the
outer disc (see Fig. \ref{fig:model}) we use a simple Keplerian velocity
profile, but this is something that will be improved on later. In a previous
paper (Naso \& Miller \cite{papI}, hereafter Paper I) we analysed the distortion
of the poloidal component of the magnetic field using a similar approach, and
found that deviations away from a dipole field can be quite significant. Here we
focus on the toroidal component and use the results of the previous model to
solve the $\phi$ component of the induction equation. We find that in general
$B_\phi$ follows a profile different from that of the analytic models, i.e. Eq.
(\ref{eq:bp_an}), and reduces to that only in a very particular case.

Following this introduction, in Sect. \ref{sec:MOD} we briefly describe our 
model, which is the same as that of Paper I; in Sect. \ref{sec:EQU} we recall 
the equations used (obtained from the induction equation), give expressions for 
the velocity and diffusivity profiles and outline our solution method (details 
of tests made on the code are given in an Appendix); in Sect. \ref{sec:RES} we 
present our numerical results; in Sect. \ref{sec:DIS} we comment on these, 
comparing them with those of the earlier analytic models, and develop our new 
suggestion for the $B_\phi$ profile. Sect. \ref{sec:CON} contains conclusions.

\section{Model}\label{sec:MOD}
 In this study, we are considering disc accretion by a neutron star having a 
dipolar magnetic field. The model is the same as that of Paper I. For a 
detailed description of it, see Section 2 of that paper; here we recall the 
main points.

We are assuming that the star is rotating about its magnetic axis, and that 
this axis is perpendicular to the plane of the disc; also, we assume that the 
fluid flow is steady and has axial symmetry everywhere. We use the kinematic 
approximation and do not consider any dynamo action, but turbulent diffusivity 
is included. The velocity field is not constrained to be purely azimuthal but 
is allowed to have non-zero components also in the other directions. We use 
spherical coordinates ($r$, $\theta$, $\phi$), with the origin being at the 
centre of the neutron star. Boundary conditions are imposed at the inner and 
outer radial edges of the disc ($r_{\rm in}$ is at the Alfven radius, and 
$r_{\rm out}$ is at $38 \, r_{\rm in} $), on top of the corona (taken as being 
a layer above and below the disc) and on the equatorial plane. Having the inner 
edge of the disc at the Alfven radius justifies the kinematic approximation to 
some extent, since in this configuration the magnetic pressure is smaller than 
the gas pressure within the region that we are considering, and so the effects 
of the plasma on the magnetic field should be larger than the magnetic feedback 
on the plasma flow. The ratio $h/r$ is taken to be constant, with the opening
angle being $8^\circ$ for the disc (measuring from the equatorial plane to the
top of the disc), and $10^\circ$ for the disc plus corona.

The inner disc region ($r < r_{\rm tr} \sim 2 \, r_{\rm in}$) and the corona 
are modelled with a larger value of $\eta$ than the other parts. In these 
regions, the kinematic approximation does not provide a good description of 
the system for two different reasons: in the corona this is because of the low 
density of the plasma (which therefore tends to follow the magnetic field
behaviour rather than being followed by it); in the inner region, it is because
the magnetic field intensity is still quite large - although the magnetic
pressure is smaller than the gas pressure, it is not yet negligible. Using a
larger value for $\eta$ in these regions makes the magnetic field less sensitive
to the plasma motion; a somewhat similar approach was used by Kueker et al.
(\cite{Retal00}). We recall that the present knowledge of the turbulent magnetic
diffusivity is quite poor and it is not a simple task to obtain a reliable
expression for the $\eta$ profile.

 As regards the velocity field: for $v_r$ we use the expression given for the 
``middle region'' of $\alpha$-discs by Shakura \& Sunyaev (\cite{SS73}). For 
$\Omega$ we take Keplerian rotation in the disc and corotation at the top of the
corona and at the inner edge of the disc, giving a maximum for $\Omega$ between
$r_{\rm in}$ and $r_{\rm tr}$. These different parts are smoothly connected
using error functions. Regarding $v_\theta$: we put it to zero in the disc but
near to the boundaries we are forced to have a non-zero value in order to be
consistent with the dipolar boundary conditions (as shown in Section 3.2 of
Paper I) and so we use a non-zero profile in the corona. In this way we are
including in the model an outflow from the surface of the corona, and this
is in agreement with recent hydrodynamic simulations of accretion flows (Jiao
and Wu, \cite{JW11}).

 Summarising, we divide the surroundings of the central object into four parts
(see Fig. \ref{fig:model}, which is repeated from Paper I): (1) the inner disc,
extending from $r_{\rm in}$ out to a transition radius $r_{\rm tr} \sim 2 \,
r_{\rm in}$ (where the diffusivity changes); (2) the outer disc, extending from
$r_{\rm tr}$ to an outer radius $r_{\rm out} = 38 \, r_{\rm in}$; (3) a corona,
above and below the disc; and (4) everything else, which we take here to be
vacuum. As a unit for radial distances, we use the Schwarzschild radius $r_{{\rm
g}}$. Within the outer disc, we focus on what we call the main region, extending
from $r_{\rm tr}$ out to the light cylinder at $r_{\rm lc} \sim 11 \, r_{\rm
in}$.


\section{Equations}\label{sec:EQU}
 In the kinematic approximation, one assumes that the velocity field remains 
fixed as specified, and the interaction between the magnetic field and the 
plasma is then described by the induction equation alone. In the presence of 
turbulence, it is more convenient to write this equation for mean fields rather 
than for the actual fields (which contain fluctuating parts as well).

The time dependence of the mean field is given by
 \begin{equation}
\partial_t\mathbf{B}=\nabla\times\left(  \mathbf{v}\times\mathbf{B} + 
\mathbf{\mathcal{E}} - \eta_{\rm \mbox{\tiny Ohm}}\nabla\times\mathbf{B}\right) 
\, {\rm ,}
\end{equation}
where $\eta_{\rm \mbox{\tiny Ohm}}=c^2/4\pi\sigma$ is the Ohmic diffusivity and
$\mathbf{\mathcal{E}}$ is the turbulent electromotive force. A common procedure
is to expand $\mathbf{\mathcal{E}}$ in terms of the mean field and its
derivatives and within the first order smoothing approximation one has
$\mathbf{\mathcal{E}} = \alpha_{\rm \mbox{\tiny{T}}} \mathbf{B}-\eta_{\rm
\mbox{\tiny{T}}} \nabla \times \mathbf{B}$, where the $\alpha_{\rm \mbox{\tiny
T}} \mathbf{B}$ term generates the so-called $\alpha$-effect. As in Paper I, we
are neglecting this effect here and the induction equation then reduces to
\begin{equation}
\partial_t\mathbf{B} = \nabla \times \left(  \mathbf{v} \times \mathbf{B}  - 
\eta \nabla \times \mathbf{B} \right) \, {\rm ,}
\label{eq:IND}
\end{equation}
where $\eta = \eta_{\rm \mbox{\tiny Ohm}} + \eta_{\rm {\mbox{\tiny T}}}$ and is
$\sim\eta_{\rm {\mbox{\tiny T}}}$, because the turbulent diffusivity is much
stronger than the Ohmic one.

 We note that the effects of a dynamo action on the disc structure have recently
been studied by Tessema \& Torkelsson (\cite{TT10I} and \cite{TT10II}), who
estimated the toroidal magnetic field generated by the dynamo to be about an
order of magnitude larger than the $B_\phi$ calculated according to the early
models. Here we show that the profile of the toroidal field can be very
different from the one suggested by those models, if the poloidal component is
not forced to be a dipole, but is instead calculated self-consistently.

As described in Paper I, our strategy consists of writing Eq. (\ref{eq:IND}) in
spherical coordinates, applying the axisymmetry and stationarity assumptions
(i.e. putting $\partial_\phi[\dots] = \partial_t[\dots] = 0$) and then solving
the final equations with the velocity field and magnetic turbulence profile
given by the model. The three components of the induction equation are
 \begin{eqnarray}
 0 &=& \partial_\theta \left\{ \sin\theta \left[ v_r B_\theta - v_\theta 
B_r - \frac{\eta}{r}[\partial_r(rB_\theta) - \partial_\theta B_r]\right] 
\right\} \label{eq:ind_r} \, {\rm ,} \\
 0 &=& \partial_r \left\{ r \hspace{0.5cm}\left[ v_r B_\theta - v_\theta 
B_r - \frac{\eta}{r}[\partial_r(rB_\theta) - \partial_\theta B_r] \right] 
\right\}\label{eq:ind_t} \, {\rm ,} \\
 0 &=& \partial_r \left\{ r \left[ v_\phi B_r - v_rB_\phi + \frac{\eta}{r} 
\partial_r(rB_\phi) \right] \right\} -\nonumber\\
&&\partial_\theta \left\{ v_\theta B_\phi - v_\phi B_\theta  -  
\frac{\eta}{r\sin\theta}\partial_\theta (B_\phi \sin\theta) \right\}  \, \rm{.}
\label{eq:ind_p}
\end{eqnarray}
 The first two equations contain only poloidal quantities and have been solved
in Paper I (making use of the magnetic stream function). Here we focus on the
third equation and solve it using the results for $B_r$ and $B_\theta$ from the
previous analysis.

We rewrite Eq. (\ref{eq:ind_p}) in the following dimensionless way:
\begin{eqnarray}
\label{eq:B_short}
\partial_x^2 B_\phi + {\rm a_{\theta\theta}}\, \partial_\theta^2 B_\phi + {\rm
a_x}\,
\partial_x B_\phi + {\rm a_\theta}\, \partial_\theta B_\phi + {\rm a_1} \,
B_\phi +
{\rm a_0} = 0 \, {\rm ,}
\end{eqnarray}
 where $x=r/r_g$ (not to be confused with the Cartesian coordinate $x$ used for
the plots), $B_\phi$ is the toroidal field in units of a reference field (as
described in Sect. \ref{sec:sol_met} below) and the dimensionless $a$
coefficients are
 \begin{eqnarray}
\label{eq:att}
\rm{a_{\theta\theta}} & = & \frac{1}{x^2} \\
\label{eq:at}
\rm{a_\theta}  & = & \frac{\partial_\theta \eta}{x^2\eta} + \frac{1}{x^2}
\frac{\cos\theta}{\sin\theta} - \frac{v_\theta r_g}{x\eta}  = \frac{1}{x} \left[
\frac{\partial_\theta \eta}{x\eta} + \frac{1}{x} \frac{\cos\theta}{\sin\theta} -
\frac{v_\theta r_g}{\eta} \right]\\
\label{eq:ax}
\rm{a_x}  & = & \frac{ \partial_x \eta}{\eta} + \frac{2}{x} - \frac{v_r
r_g}{\eta}  \\
\nonumber
\rm{a_1}  & = & \frac{ \partial_x \eta }{x\eta} - \frac{
r_g \partial_x (xv_r)} {x\eta} +
\frac{\partial_\theta\eta} {x^2\eta}\frac{\cos\theta}{\sin\theta} -\\
\nonumber
&&\frac{1}{x^2} \frac{\cos^2\theta}{\sin^2\theta} - \frac{r_g \partial_\theta
v_\theta}{x\eta} -\frac{1}{x^2} = \\
\label{eq:a1}
& = & \frac{1}{x} \left[ \frac{ \partial_x \eta }{\eta} - \frac{
r_g\partial_x(xv_r)} {\eta} +
\frac{\partial_\theta\eta} {x\eta}\frac{\cos\theta}{\sin\theta} - \right.\\
\nonumber
&& \left. \frac{1}{x} \frac{\cos^2\theta}{\sin^2\theta} - \frac{r_g
\partial_\theta v_\theta}{\eta} -\frac{1}{x} \right] \\
\nonumber
\rm{a_0}  & = & \frac{r_g\partial_x(x v_\phi B_r)}{x\eta} + \frac{r_g
\partial_\theta(v_\phi B_\theta)}{x \eta} = \\
\label{eq:a0}
&& = \frac{r_g}{x\eta} \left[ \partial_x(x v_\phi B_r) + \partial_\theta(v_\phi
B_\theta)  \right].
\end{eqnarray}

All of these coefficients have direct analytic expressions, except for the 
last one ($a_0$) which contains $B_r$ and $B_\theta$, whose values are taken
from the previous numerical calculations. Note that the magnetic field
components appearing in Eq. (\ref{eq:a0}) are dimensionless.

\subsection{Poloidal velocity and diffusivity}\label{sec:VELETA}
 The poloidal components of the velocity field and the diffusivity have already
been discussed in Paper I. Here we use the same profiles as before; for the sake
of clarity, we give the expressions again below.

For $v_r$, we use
\begin{eqnarray}
\label{eq:vr}
\nonumber
v_r(x)&=&2 \times 10^6 \, \alpha^{4/5} \, \dot{m}^{2/5} \, m^{-1/5}\\
&& \times \, (3/x)^{2/5} \left[ 1-(3/x)^{0.5} \right]^{-3/5}
\hspace{0.5cm}[\rm{cm}\,\rm{s}^{-1}] \, {\rm ,}
\end{eqnarray}
 where $m$ is the stellar mass in solar mass units, $\dot{m}$ is the accretion 
flux in units of the critical Eddington rate and $\alpha$ is the standard 
Shakura-Sunyaev viscosity coefficient. Using typical values ($\alpha=0.1$, 
$\dot{m} = 0.03$ and $m = 1.4$) this gives
 \begin{equation}
 v_r(x)\approx7.3\,10^4 \cdot (3/x)^{2/5} \left[ 1-(3/x)^{0.5} \right]^{-3/5} 
\rm{cm}\,\rm{s}^{-1}\, \rm{.}
\end{equation}
 For $v_\theta$, we use
\begin{eqnarray}
\label{eq:vt}
v_\theta(r,\theta) &=& \left\{ \begin{array}{ll}
0 & \textrm{ \hspace{.25cm} in the disc} \\
\frac{1}{2}\,v_r\,\tan\theta & \textrm{ \hspace{.25cm} in the corona}
\end{array} \right. 
\end{eqnarray}
where the transition in $v_\theta$ between the two regions is made using
 \begin{equation}
v_\theta(r,\theta) = f_1(\theta)\,\frac{1}{2}\,v_r\,\tan\theta  
\end{equation}
 with
 \begin{equation}
 f_1(\theta) = \frac{1}{2} \left[  1 + \rm{erf}\left(\frac{- \theta + 
\theta_D} {d} \right) \right] \, {\rm ,}
\end{equation}
 where $\theta_D$ is at the upper surface of the disc (i.e. $82^\circ$)
and $d = 10^{-3}$ radians (i.e. $0.057^\circ$). 

For the diffusivity, we use
\begin{eqnarray}
\label{eq:eta}
  \eta(r,\theta) &=& \eta_0 \, \left\{1 + \big[
\eta_\theta(\theta) + \eta_r(r) \big] \, \left[ \frac{\eta_{\rm c}}{\eta_0}
- 1 \right] \right\} \, {\rm ,}
\end{eqnarray}
 where $\eta_0$ and $\eta_{\rm c}$ are the values in the  main disc region and
the corona (see Fig. \ref{fig:model}), for which we use the values $10^{10}$
cm$^2$ s$^{-1}$ and $10^{12}$ cm$^2$ s$^{-1}$ respectively (we also use other
values to study the impact on the results of varying $\eta$). For $\eta_\theta
(\theta)$ and $\eta_r(r)$ we use joining functions of the form
 \begin{equation}
 f(x) = \frac{1}{2}\left[1 + {\rm erf}\left(\frac{- x + x_c} {d} \right) 
\right] \, {\rm ,}
\end{equation}
 with $x_c=\theta_D$ and $r_{\rm tr}$ for $\eta_\theta$ and $\eta_r$,
respectively, and with $d$ being the width of the transition in the error 
function, for which we use $d=5\,r_g$ in the radial direction and $d = 2^\circ$ 
in the $\theta$ direction\footnote{Note that the error-function widths in the 
radial and angular directions are larger than those used in Paper I (there $d_r 
= 2\,r_{\rm g}$ and $d_\theta=10^{-3}$ radians $ = 0.057^\circ$). The reasons 
for this are explained in Appendix \ref{app:test_mod}.}.

\subsection{Azimuthal velocity}
 We are modelling the main disc region as rotating with Keplerian velocity, with 
the corona being taken as a transition layer where the velocity goes from 
Keplerian to corotation. In the inner region, we join the Keplerian flow to
corotation at the radial inner edge, again using an error function. We recall
that in Paper I we showed that a strictly dipolar field, without distortions,
can in principle be a stationary solution of the induction equation provided
that the velocity field fulfils the two conditions:
 \begin{eqnarray}
\label{eq:vt_an_const}
v_\theta &=& \frac{1}{2} \, \tan\theta \, v_r\\
\nonumber
{\rm and}\\
\label{eq:omega_an_const}
\Omega &\propto& r^{-\gamma/2} \, \sin^{\gamma}\theta \, \rm{.}
\end{eqnarray}
 From Eq. (\ref{eq:omega_an_const}) one sees that corotation, which is obtained 
by choosing $\gamma=0$, is consistent with having dipolar conditions (which is 
what we are using here).

For the magnetic field intensities and neutron star spin rates which we are
using as standards ($B \sim 10^8$ G and $P \sim 10$ ms), the corotation point is
outward of the inner edge of the disc (which is the standard condition required
for being in the accretion regime) and therefore $\Omega$ should reach a maximum
at some location and then decrease again, as one moves inwards. Summarising, we
use the following profile:
\begin{equation}
 \label{eq:omega}
 \Omega(r,\theta) = \left\{
 \begin{array}{cl}
  \Omega_K(r) & \hspace{0.3 cm} \mbox{in the main region:}\\
& \hspace{0.3 cm} \theta\in[\theta_D,\pi/2], r\in[r_{\rm tr},
r_{lc}]\\
  & \\
  \mbox{smooth join in $\theta$} & \hspace{0.3 cm} \mbox{in the corona:}\\
 & \hspace{0.3 cm} \theta\in[\theta_C, \theta_D]\\
  & \\
  \mbox{smooth join in $r$} & \hspace{0.3 cm} \mbox{in the inner disc:}\\
 & \hspace{0.3 cm} r\in[r_{\rm in}, r_{\rm tr}]\\
  & \\
  \Omega_{\rm s} & \hspace{0.3 cm} \mbox{at ghost zones: } \theta = \theta_C -
\Delta \theta\\
& \hspace{0.3 cm} \mbox{at inner edge: } r = r_{\rm in}
 \end{array}
 \right.
\end{equation}
 where $\theta_C$ is the upper surface of the corona, $\Delta \theta$ is the
angular resolution, $\Omega_{\rm s}$ is the stellar spin rate, $\Omega_K$ is the
Keplerian angular velocity $\sqrt{GM/r^3}$ and the two smooth connections are
made using the error functions given in Eqs. (\ref{eq:erf_t}) and
(\ref{eq:erf_r}) below.

\begin{figure}[ht]
 \centering
 \includegraphics[width=0.4\textwidth]{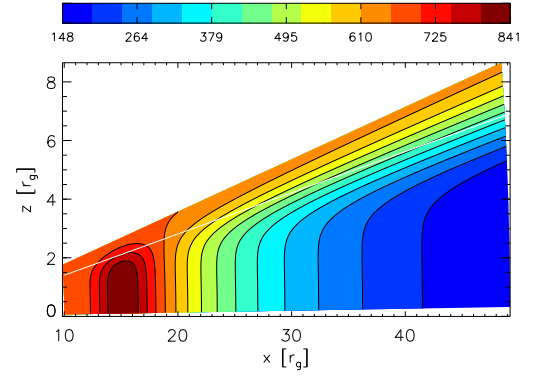}
 \caption{Contour plot of $\Omega(r,\theta)$. This is the same for all of the
configurations. The straight white line indicates the boundary between the disc
and the corona. The corotation point is at $r = 18.8 \, r_{\rm g}$.}
 \label{fig:omega}
\end{figure}

As regards the smooth joins, in the $\theta$ direction we write
 \begin{eqnarray}
\nonumber
 \tilde{\Omega}(r,\theta) &=& \Omega_K(r) \, f_1(\theta) + \Omega_{\rm s} \,
[1-f_1(\theta)]\\
 &=&[ \Omega_K(r) - \Omega_{\rm s}] \, f_1(\theta) + \Omega_{\rm s}
\end{eqnarray}
where:
\begin{equation}
\label{eq:erf_t}
 f_1(\theta) = \frac{1}{2}\left[ 1 + {\rm erf} \left(
\frac{\theta-\theta_D}{d} \right) \right]
\end{equation}
 with $d=10^{-2}$ radians (i.e. $0.57^\circ$), and we then have a modification 
in the radial direction giving
 \begin{eqnarray}
 \nonumber
 \Omega(r,\theta) &=& \tilde{\Omega}(r,\theta)\,f_2(r)+\Omega_{\rm
s}\,[1-f_2(r)]\\
 &=&[ \tilde{\Omega}(r,\theta) - \Omega_{\rm s}] \, f_2(r) + \Omega_{\rm s} \, 
{\rm ,}
\end{eqnarray}
where
\begin{equation}
\label{eq:erf_r}
 f_2(r) = \frac{1}{2}\left[ 1 + {\rm erf} \left( \frac{x-x_c}{d} \right)
\right]
\end{equation}
 with $d = 2 \, r_g$. See Fig. \ref{fig:omega} for a contour plot of
$\Omega(r,\theta)$ (made in terms of Cartesian coordinates $x$ and $z$.)

\subsection{Boundary conditions for $B_\phi$}
 In our model we take the region outside of the disc and its associated corona
to be vacuum, and suppose that there is no toroidal component of the magnetic
field there (i.e. $B_\phi = 0$). We impose the same condition also at the
equatorial plane because $B_\phi$ has to be antisymmetric across it.

When $B_\phi = 0$, Eq. (\ref{eq:ind_p}) reduces to $a_0=0$, i.e.
\begin{equation}
 \label{eq:bc}
 \partial_x(x v_\phi B_r) + \partial_\theta(v_\phi B_\theta) = 0
\end{equation}
 at the top of the corona, at the inner edge of the disc and on the equatorial 
plane. We need our choice of the boundary conditions (i.e. purely dipolar field 
and corotation) to represent a solution of this equation. In Paper I, we showed 
that corotation is consistent with a pure dipolar field, as already mentioned 
in the previous subsection (see Eq. (\ref{eq:omega_an_const})). In addition we 
note here that, on the equatorial plane, Eq. (\ref{eq:bc}) is trivially
satisfied because $B_r = 0$ and both $v_\phi$ and $B_\theta$ are symmetric with 
respect to this plane.

\subsection{Solution method}\label{sec:sol_met}
 In order to get to Eq. (\ref{eq:B_short}) we first expanded out all of the
derivatives present in Eq. (\ref{eq:ind_p}) and then put the result into a
dimensionless form using $x=r/r_{{\rm g}}$ as the radial coordinate and 
measuring $B_\phi$ in units of $B_\phi^0$, which is a reference value for the
magnetic field, taken to be $r_0^2\,B_0$, where $B_0$ is the magnitude of the
dipolar field at the stellar equator and $r_0$ is the dimensionless neutron-star
radius. As in Paper I, we use canonical values for the mass and radius of the
neutron star, $1.4\,M_\odot$ and 10 km respectively, and take $B_0 = 3 \times
10^8$ G, as typical for a millisecond pulsar.

Equation (\ref{eq:B_short}) is a non-homogeneous elliptic partial differential 
equation for $B_\phi$ and we have solved it using the Gauss-Seidel relaxation 
method which uses a finite-difference technique, approximating the operators by 
discretizing the functions over a grid. At any given iteration step, the values 
of $B_\phi$ at the various grid points are written in terms of values at the 
previous step, or at the present step in the case of locations where it has 
already been updated. Details of the numerical scheme are given in the Appendix 
of Paper I, where we used the same method to solve the elliptic partial 
differential equation for the magnetic stream function.

 Before applying the numerical scheme to the actual problem that we want to 
solve, we performed a series of tests on the code, which are described in 
detail in the Appendix. We used several configurations, with many different 
numbers of grid points, profiles of the turbulent diffusivity, initial 
estimates for the toroidal field, locations for the outer radial boundary of 
the grid and values for the iteration time step. In this way we have checked 
the code stability and convergence, have optimised the iteration procedure and 
have determined where best to place the outer radial boundary of the grid (so 
that the outer boundary conditions do not significantly influence the solution 
in our region of interest).

 All of the results presented in this paper have been obtained using a grid 
spacing of $\Delta r = 0.74 \, r_g$ and $\Delta \theta = 0.125^\circ $. The 
final maximum residual was of the order of $10^{-13}$ and saturation was reached
after about $4 \times 10^6$ iterations, the iteration time step being $4.46
\times 10^{-4}$ (about $95\%$ of the critical one, beyond which the method gives
a divergent solution). As a first estimate for the profile of the magnetic field
we have used a Gaussian. The numerical domain was $x \in [10, 380]$, $\theta \in
[80, 90]$ (in degrees), which we covered with a homogeneous grid of $501 \times
81$ points. The profile of the poloidal field (which is present in the
expression for the coefficient $a_0$) was calculated by running the code used in
Paper I with the same profile of $\eta$ and $\mathbf{v}$ and the same
resolutions as used in this analysis. However, since for the poloidal
calculation there is a stronger dependence on the outer radial boundary
conditions, we have placed $r_{\rm out}$ at $750\,r_{\rm g}$ and used a $1001
\times 81$ grid. In general the number of radial points used in the poloidal
calculation ($N_i^{\rm pol}$) and in the toroidal one ($N_i^{\rm tor}$) are
related through the following condition (which comes from equating the spatial
resolutions):
 \begin{equation}
\label{eq:Ni}
N_i^{\rm tor} - 1 =  \frac{r_{\rm out}^{\rm tor} - r_{\rm in}} {r_{\rm out}^{\rm
pol} - r_{\rm in}} (N_i^{\rm pol} -1) \, \rm{.}
\end{equation}
 For the poloidal calculation the final maximum residual was of the order of 
$10^{-9}$ and was reached after about $7 \times 10^7$ iterations. The iteration 
time step was the same as for the toroidal calculation.

\section{Results}\label{sec:RES}
\subsection{Reference configuration}
\begin{figure}[ht]
 \centering
 \includegraphics[width=0.4\textwidth]{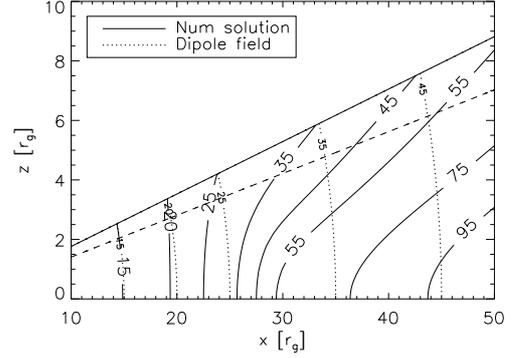}
 \caption{Poloidal field lines from the numerical solution (solid) and those 
for a dipole (dotted). In this configuration, $v_0=10^5$ cm s$^{-1}$ while the
diffusivity $\eta_0=10^{10}$ cm$^2$ s$^{-1}$ with a transition width in $\theta$
direction of $2^\circ$ ($3.5 \times 10^{-2}$ radians). Straight lines indicate
the top surface of the corona (solid) and the boundary between disc and corona
(dashed).}
	\label{fig:Bref_lines}
\end{figure}

\begin{figure}[ht]
	\centering
	\includegraphics[width=0.4\textwidth]{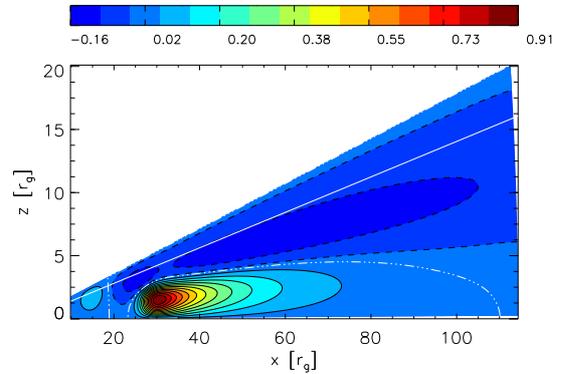}
	\caption{Contour plot of the toroidal field in Cartesian coordinates, in
the region $r = [10-115] \, r_g$ and $\theta = 80^\circ - 90^\circ$. The black
solid lines are lines of positive $B_\phi$, while black dashed lines are used
for negative $B_\phi$. $B_\phi=0$ is shown with triple dotted-dashed white 
lines. The straight white line indicates the boundary between disc and corona.}
	\label{fig:ref_cont_10.115}
\end{figure}

 As mentioned in Sect. \ref{sec:EQU}, we have used a slightly different 
configuration from that considered in Paper I (the transition in the $\eta$ 
profile is wider and the resolution in the angular direction is increased). 
The poloidal field for the new configuration, resulting from solving Eqs. 
(\ref{eq:ind_r}) and (\ref{eq:ind_t}), is shown in Fig. \ref{fig:Bref_lines}. 
If we additionally assume a profile for the angular velocity, we can then solve
Eq. (\ref{eq:B_short}). The result, for the profile given by Eq.
(\ref{eq:omega}), is shown in Fig. \ref{fig:ref_cont_10.115} as a contour 
plot of $B_\phi$, for all of the region of interest (i.e. for $r \in [10, 115]
\, r_g$ and including the corona). Contours for positive values of $B_\phi$,
representing toroidal field lines rotating in the same direction as the neutron
star, are shown with black solid lines; while those for negative values of
$B_\phi$ (rotating in the opposite direction) are shown with black dashed lines.
Triple dotted-dashed white lines show where $B_\phi=0$.

\begin{figure}[hb!]
	\centering
	\includegraphics[width=0.4\textwidth]{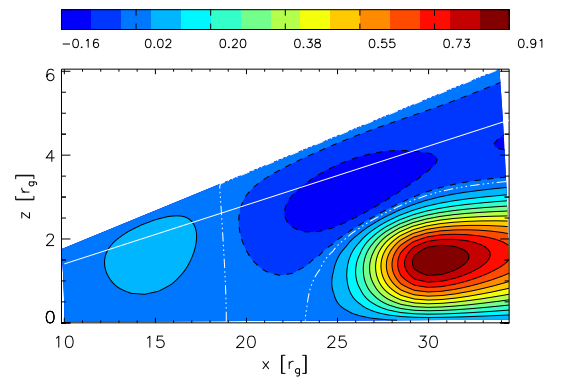}
	\caption{The same as in Fig. \ref{fig:ref_cont_10.115} but showing an
expanded view of the region $r = [10-35] \, r_g$. The corotation point is at 
$r = 18.8 \, r_{\rm g}$. }
	\label{fig:ref_cont_11.33}
\end{figure}

The toroidal field shows a quite structured profile, with two (positive) maxima 
and two (negative) minima. Surprisingly, the global maximum is located outward
of the corotation point and is positive, in contrast to the standard picture
according to which the sign of $B_\phi$ is the same as that of the relative
angular velocity between the disc matter and the star. There is also a quite
striking vertical structure, the global minimum being located just above and to
the left of the global maximum. As in the analytic models, $B_\phi$ becomes zero
near to the corotation point ($r_{\rm cor} = 18.8 \, r_{\rm g}$), but here one sees that for almost every value of $r>r_{\rm cor}$ there is a value of $\theta$ where the field
is zero, so that there are places with zero $B_\phi$ throughout all of the main
disc region.

In Fig. \ref{fig:ref_cont_11.33} we show the $B_\phi$ contour plot in the region 
$r \in [10, 35] \, r_{\rm g}$, so that the structure of the magnetic field in 
this part can be seen more clearly. In Table \ref{tab:min_max} we give the 
coordinates of the minima and maxima and the magnitude of the toroidal field at 
those locations, measured in units of the stellar field strength $B_0$ (which is 
taken to be $3 \times 10^8$ G). All of these main features are located in the 
region $r \in [10, 55] \, r_g$; the remainder of the disc can be divided into two 
zones: one where $B_\phi>0$ (in Figs. \ref{fig:ref_cont_10.115} and 
\ref{fig:ref_cont_11.33} this is below the long white triple dotted-dashed curve 
that crosses the equatorial plane at $x=23$ and $x=110$), and the other where, 
instead, it is negative. The minimum latitude of the region with positive
$B_\phi$ is about $84^\circ$.

\begin{table}[ht]
 \caption{Locations of extrema of the toroidal field.}
\begin{center}
\begin{tabular}{lccc}
\hline
\hline
Extremum & $x\,[r_g]$ & $\theta$ & $B_\phi[B_0]$ \\
\hline
Global maximum & $30.5$ & $87.1^\circ$ & $0.78$  \\ 
Local maximum  & $14.5$ & $83.1^\circ$ & $0.092$ \\ 
Global minimum & $25$   & $83.4^\circ$ & $-0.17$ \\ 
Local minimum  & $52$   & $83.6^\circ$ & $-0.13$ \\ 
\hline
\end{tabular}
 \tablefoot{ The last column reports the intensity of the toroidal magnetic 
field at the corresponding locations, measured in units of the stellar field 
strength $B_0$ (which is taken to be $3 \times 10^8$ G).}
 \label{tab:min_max}
\end{center}
\end{table}
\vspace{-0.5cm}

\begin{figure}[ht]
 \centering
 \includegraphics[width=0.4\textwidth]{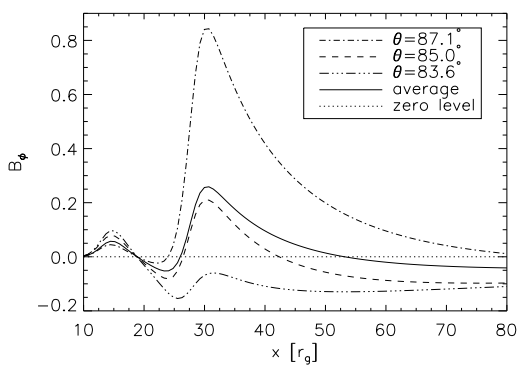}
 \caption{Toroidal field strength plotted against $r$ at different values of
$\theta$, in the region $r = [10-50] \, r_g$ (it is equal to zero on the
equatorial plane). Negative values mean that $B_\phi$ is pointing backwards
with respect to the disc rotation. The thick solid line is the shell average of 
$B_\phi$.}
 \label{fig:multiple_theta_and_avg}
\end{figure}

Radial profiles of $B_\phi$ are shown in Fig. 
\ref{fig:multiple_theta_and_avg} for several values of the latitude and for the 
shell average. We show the profiles for $\theta = 87.1^\circ$ (where the global
maximum is), $\theta = 83.6^\circ$ (where the local minimum is, and which is
very close also to the local maximum and the global minimum) and for an
intermediate value $\theta = 85^\circ$. The curves all pass through zero at 
the corotation point ($r = 18.8\, r_g$), at least to within the accuracy of the 
calculation. The large positive $B_\phi$ peak at about $30\,r_g$ is
progressively reduced as one moves from the mid plane of the disc towards the 
corona, and eventually becomes a negative local maximum. One can calculate the 
shell average of the radial profile, i.e. an average of $B_\phi$ over $\theta$ 
for each $r$, and this is also shown in Fig. \ref{fig:multiple_theta_and_avg} as
a thick solid line. This average reproduces quite well the general behaviour of
the toroidal field and shows all of its main features.

The $\theta$ dependence of the sign of $B_\phi$ is a key result, that can have 
quite dramatic consequences for the calculation of the magnetic torque exerted 
on the neutron star. In fact, it is usually assumed that the torque depends 
only on $r$, being positive inside the corotation radius and negative outside 
it (see Eqs. (\ref{eq:bp_an}) and (\ref{eq:tor})), whereas we now see regions 
of positive $B_\phi$ even outward of the corotation point. Therefore we need to
rethink the discussion of which regions in the disc tend to spin the star up or
down. An appropriate approach for calculating the torque requires integration in
both directions: we plan to perform this analysis in a future work.

Finally, we note here that the magnitude of the toroidal component of the 
magnetic field is typically larger than that of the poloidal component. For 
example, the maximum of the shell average of the poloidal component $\left(
\left< \sqrt{B_r^2 + B_\theta^2} \, \right> _\theta \right)$ is $\sim 3.6 \times
10^{-3}$, while the maximum of $\left<B_\phi\right>_\theta$ is $\sim 2.3 \times
10^{-1}$. In terms of energy conservation, one has to bear in mind that any
change in the magnetic energy has to be compensated by a corresponding opposite
change in the plasma energy, while in the present model we are taking the flow
pattern to be fixed. It is important for the back-reaction to be consistently
taken into account and this will be done in subsequent work. Also, the
distortions of the toroidal field will be limited by magnetic reconnection.

\vspace{-0.2cm}

\subsection{Configurations with larger $\eta_0$}
 We have already mentioned in Sect.~\ref{sec:MOD} that there are big 
uncertainties about how to model the turbulent magnetic diffusivity within the 
disc and the surrounding corona. This quantity is often discussed in terms of
the turbulent magnetic Prandtl number, $P_{\rm m} \equiv \nu/\eta$, which links
it with the turbulent viscosity.

Within the kinematic approximation, one does not solve self-consistently for 
the velocity field and so it is necessary assume some profile for it. As 
outlined in Sect. \ref{sec:VELETA}, in the present calculations we are using 
the velocity prescription given by the Shakura $\&$ Sunyaev thin disc model,
which also embodies a particular connection between the viscosity and other
disc quantities. This gives (following e.g. Szuszkiewicz $\&$ Miller
\cite{SM2001}) $\nu = \alpha \, h \, c_{\rm s}$, with $\Omega_k^2 \, h^2 = 6 \,
p/\rho$ (assuming vertical hydrostatic equilibrium), where $c_{\rm s}$ is the
sound speed, $p$ is the pressure and $\rho$ is the density. Using the isothermal
sound speed, the Keplerian angular velocity profile and the parameters given in
Sect. $\ref{sec:VELETA}$, one gets $\nu \sim 10^{13}$ cm$^2$ s$^{-1}$ which, in
turn, gives $P_{\rm m} \sim 10^3$ in the disc and $\sim 10$ in the corona of
our reference model. These values may seem rather high, but one should be
cautious about using them because there are several further factors which need
to be taken into account.

Note, first, that calculating the viscosity with the $\alpha$ disc model 
certainly gives an over-estimate for $P_{\rm m}$, because one is neglecting the 
contribution of the magnetic field in the equation of motion. Also, there is a 
degeneracy in the profile of the velocity field, in the sense that one can 
obtain the same velocity profile, and hence the same results for our numerical 
calculations, using different combinations of $\alpha$ and $\dot{m}$: an 
increase/decrease by a factor of $\lambda$ in the accretion rate, together with 
a decrease/increase by a factor of $\sqrt{\lambda}$ in $\alpha$ (and hence 
$\nu$), causes no change in the velocity profile. These considerations can 
easily bring down the true values of $P_{\rm m}$ for our calculations well 
below the approximations quoted above. In any case, there is currently no 
general consensus about the correct value for $P_{\rm m}$: 
we note that rather high values have recently been found in some numerical 
MRI simulations (e.g. see Takahashi $\&$ Masada \cite{TM2011}; Romanova et al.
\cite{Retal2011}).

\begin{figure}[ht]
\centering
\subfloat[$\eta_0 = 4 \times 10^{10}$ cm$^2$ s$^{-1}$]{%
\label{fig:ref_eta_pol_a}
\includegraphics[width=0.4\textwidth]{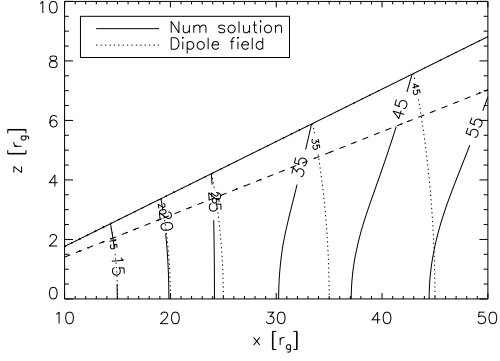} 
}\\
\subfloat[$\eta_0 = 10^{11}$ cm$^2$ s$^{-1}$]{%
\includegraphics[width=0.4\textwidth]{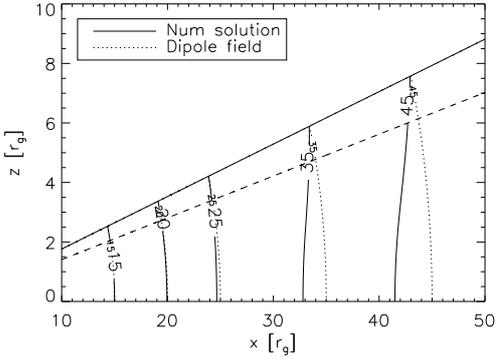}
\label{fig:ref_eta_pol_b}
}
 \caption{Poloidal magnetic field lines. Comparison between two configurations
having values of $\eta_0$ larger than that of the reference configuration.}
\label{fig:ref_eta_pol}
\end{figure}

We chose our values for $\eta$ bearing in mind which values would give 
significant field distortions in the disc and therefore be most interesting.  
However, it is clearly important to investigate the effects of varying these 
numbers, and so we also made some calculations using larger values of $\eta$ 
(smaller $P_{\rm m}$). We show here results for $\eta_0 = 4 \times 10^{10}$ 
cm$^2$ s$^{-1}$ and $\eta_0 = 10^{11}$ cm$^2$ s$^{-1}$ with $\eta_{\rm c}$ being, 
as usual, two orders of magnitude larger (in the reference configuration we used 
$\eta_0 = 10^{10}$ cm$^2$ s$^{-1}$).

\begin{figure}[ht]
\centering
\subfloat[$\eta_0 = 4 \times 10^{10}$ cm$^2$ s$^{-1}$]{%
\label{fig:ref_eta_B3_a}
\includegraphics[width=0.4\textwidth]{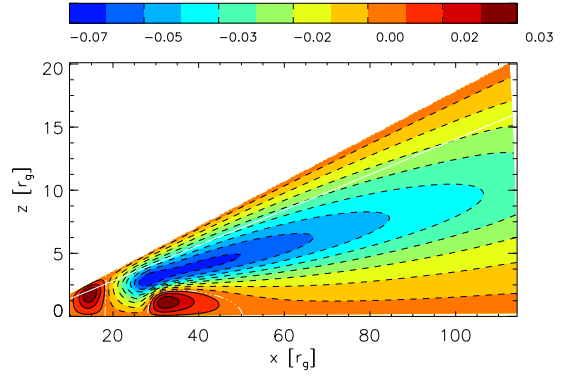}
}\\
\subfloat[$\eta_0 = 10^{11}$ cm$^2$ s$^{-1}$]{%
\label{fig:ref_eta_B3_b}
\includegraphics[width=0.4\textwidth]{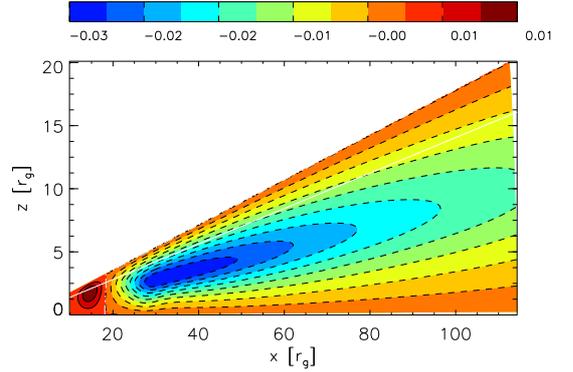}
}
 \caption{Contour plots of $B_\phi$. Comparison between two configurations
having values of $\eta_0$ larger than that of the reference configuration.}
\label{fig:ref_eta_B3}
\end{figure}

\begin{figure}[ht]
\centering
\includegraphics[width=0.4\textwidth]{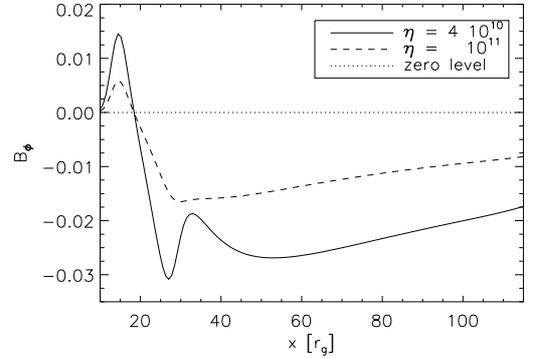} 
 \caption{Shell average of the toroidal component. Comparison between two 
configurations having turbulent magnetic diffusivity, $\eta_0$, larger than that
of the reference configuration. The values of $\eta_0$ are $4 \times 10^{10}$
cm$^2$ s$^{-1}$ for the solid curve and $10^{11}$ cm$^2$ s$^{-1}$ for the dashed
curve.}
 \label{fig:ref_eta_avgs}
\end{figure}

 Results for both configurations are presented in Figs.~\ref{fig:ref_eta_pol},
\ref{fig:ref_eta_B3} and \ref{fig:ref_eta_avgs}, where we show the poloidal
field lines, the contour plot of the toroidal field component and the radial
profile of its shell average, respectively\footnote{We have also run a
configuration with $\eta_0 = 10^{12}$ cm$^2$ s$^{-1}$ and found results
completely in agreement with the trends shown here.}.


Increasing $\eta_0$ (i.e. decreasing the magnetic distortion function $D_{\rm 
m}$, our generalisation of the magnetic Reynolds number introduced in Paper I) 
makes the field diffuse more efficiently and therefore the solution gets 
progressively nearer to being a dipole field (see Fig. \ref{fig:ref_eta_pol}). 
Reducing the poloidal field distortions in turn changes the toroidal 
component, and the modifications are quite substantial. The structure with four
extrema gradually turns into one with only two (see Fig. \ref{fig:ref_eta_B3}),
where the sign of $B_\phi$ is positive for radii smaller than the corotation
radius and negative for larger ones. This transition is clearly seen when
plotting the shell average of the toroidal field (see Fig.
\ref{fig:ref_eta_avgs}). The behaviour of the sign of the shell average of
$B_\phi$ is then the same as that for $B_\phi$ in the early analytic models,
although the details of the profiles have significant differences (see Fig.
\ref{fig:Bphi_cmp} below).

\begin{figure}[ht]
\centering
\subfloat[$\eta = 10^{11}$ cm$^2$ s$^{-1}$]{%
\label{fig:eta_const_pol_a}%
\includegraphics[width=0.4\textwidth]{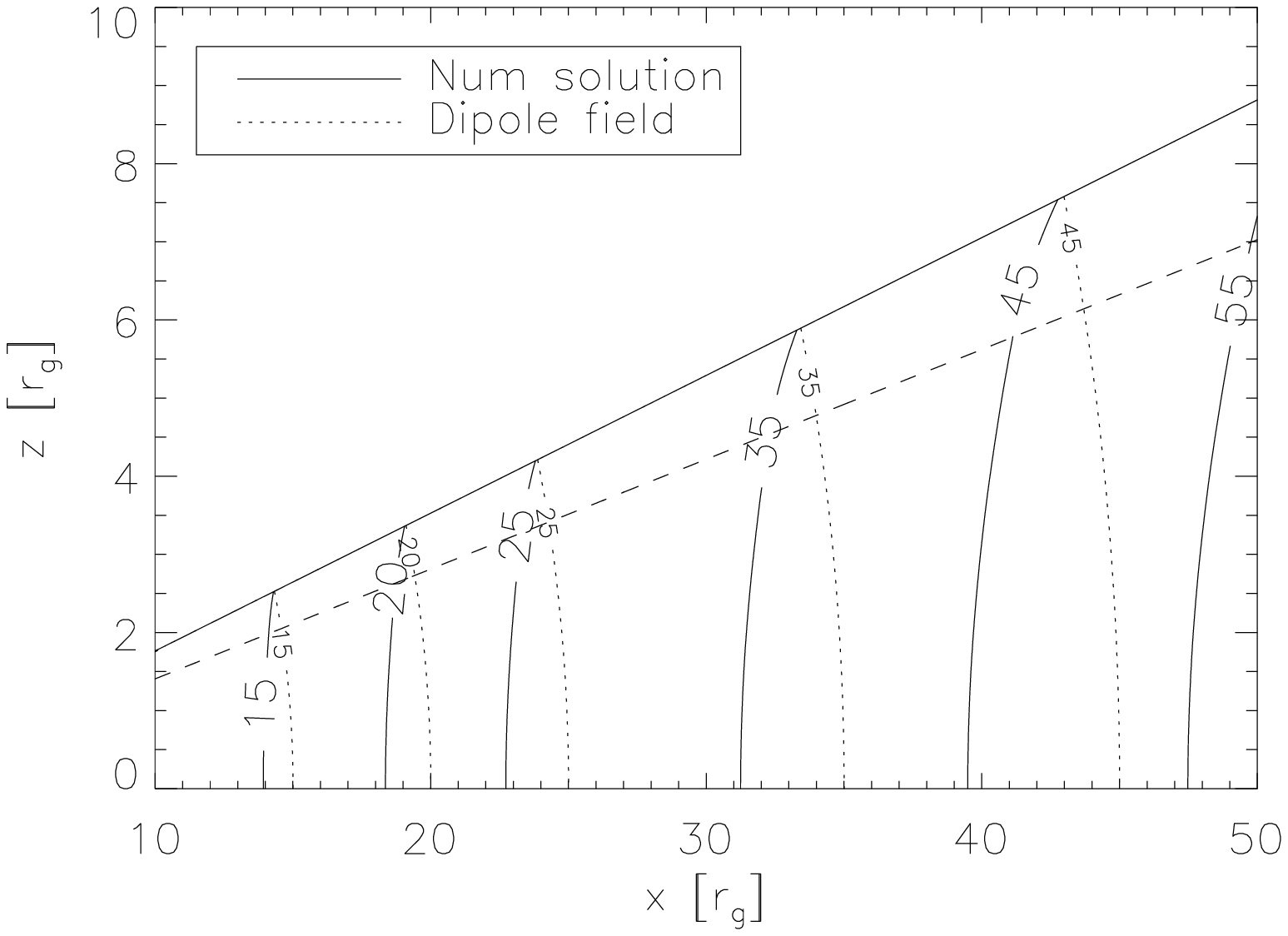}
}\\
\subfloat[$\eta = 10^{12}$ cm$^2$ s$^{-1}$]{%
\label{fig:eta_const_pol_b}%
\includegraphics[width=0.4\textwidth]{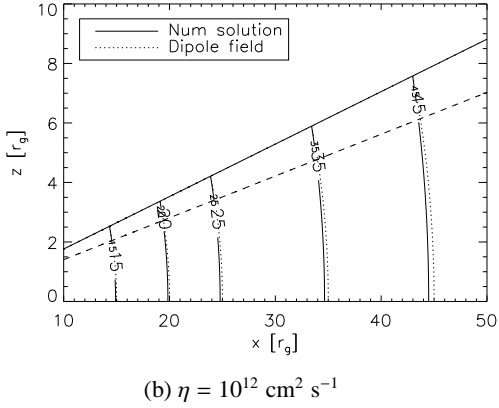}
}%
 \caption{Poloidal magnetic field lines. Comparison between two configurations
with constant $\eta$.}
\label{fig:eta_const_pol}
\end{figure}
\begin{figure}[ht]
\centering
\includegraphics[width=0.4
\textwidth]{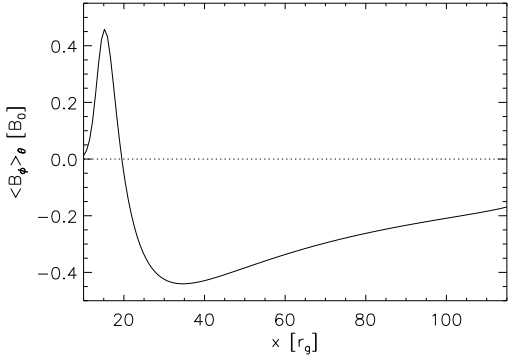}
 \caption{Shell average of the toroidal component for $\eta = 10^{11}$ cm$^2$ 
s$^{-1}$. For $\eta = 10^{12}$ cm$^2$ s$^{-1}$, the curve has exactly the same 
shape, but the values are about $100$ times smaller (in absolute value).}
 \label{fig:eta_const_av}
\end{figure}

Finally we note that the magnitude of the toroidal component decreases 
with increasing $\eta_0$, the maximum of the shell average being $\sim 14 
\times 10^{-3} \, B_0$ for the configuration with $\eta_0 = 4 \times 10^{10}$ 
cm$^2$ s$^{-1}$ and $\sim 6 \times 10^{-3} \, B_0$ for the one with $\eta_0 = 
10^{11}$ cm$^2$ s$^{-1}$ (compare the shell averages in Fig. 
\ref{fig:ref_eta_avgs} with each other and with the one in Fig. 
\ref{fig:multiple_theta_and_avg}), while the maximum for the poloidal 
component has remained at $\sim 3.5 \times 10^{-3} \, B_0$ (this is because the 
maximum for the poloidal componant occurs at the inner edge, where the field 
depends more on the boundary conditions than on the value of $\eta$).

\begin{figure}[ht]
\centering
\subfloat[$\eta = 10^{11}$ cm$^2$ s$^{-1}$]{%
\label{fig:eta_const_B3_a}
\includegraphics[width=0.4\textwidth]{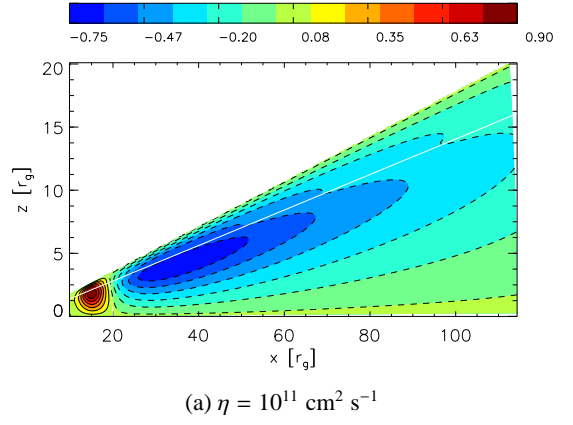}
}\\
\subfloat[$\eta = 10^{12}$ cm$^2$ s$^{-1}$]{%
\label{fig:eta_const_B3_b}
\includegraphics[width=0.4\textwidth]{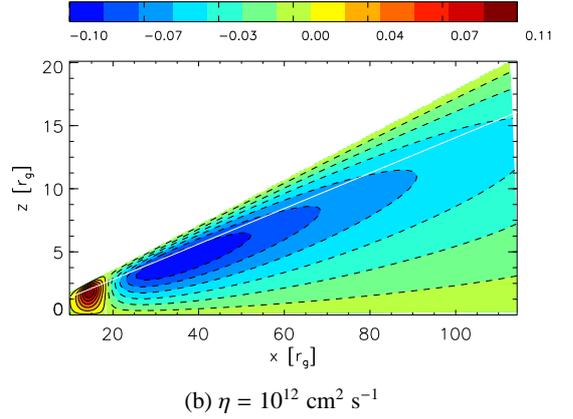}
}
 \caption{Contour plots of $B_\phi$. Comparison between two configurations
with constant $\eta$.}
\label{fig:eta_const_B3}
\end{figure}

\subsection{Configurations with constant $\eta$}
Although the $\eta$ profile that we have used so far is the one that we consider
to be the most appropriate when studying accretion within the kinematic
approximation (for the reasons given in Sect. \ref{sec:MOD}), we have considered
also configurations where the diffusivity is constant through all of the disc
and the corona, in order to have a clear understanding of how sensitive the
results are to this quantity. Cases with $\eta = 10^{10}$ cm$^2$ s$^{-1}$ and
$4\times 10^{10}$ cm$^2$ s$^{-1}$ proved unsatisfactory because of having very
abrupt changes away from the dipole field at the top of the corona (this is
exactly the behaviour that we have tried to avoid by using a larger value of
$\eta$ in the corona in our reference configuration). Results for $\eta =
10^{11}$ cm$^2$ s$^{-1}$ and $\eta = 10^{12}$ cm$^2$ s$^{-1}$ are shown in
Figs.~\ref{fig:eta_const_pol}, \ref{fig:eta_const_av} and
\ref{fig:eta_const_B3}.

For configurations with constant $\eta$, the analysis is made simpler since 
there are no regions with diffusivity gradients. As stated in the previous 
subsection, using a larger value of $\eta$ reduces the deviations away from the 
dipolar field. However, in contrast with the previous case, the distortions are
now more uniform throughout the disc (compare Figs.~\ref{fig:ref_eta_pol} and
\ref{fig:eta_const_pol}), because $\eta$ is constant and the magnetic distortion
function is monotonically decreasing with $r$ (while previously $\partial_r
D_{\rm m}$ had a peak). We note that deviations away from a pure dipole are very
small when using $\eta=10^{12}$ cm$^2$ s$^{-1}$, and for $\eta$ comparable to or
larger than this, the poloidal component of the field can be well approximated
by a dipole. As regards the toroidal component, it has only two extrema which
are both located just beneath the surface of the disc (see
Figs.~\ref{fig:eta_const_av} and \ref{fig:eta_const_B3}). These two extrema have
been observed in all of the configurations which we have studied; they are also
present in the models of Wang (\cite{W87}) and Campbell (\cite{C87}, \cite{C92})
and therefore seem to be robust features.

\section{Discussion}\label{sec:DIS}

\subsection{Comparison with analytic models} 
 In the analytic models of Wang (\cite{W87}) and Campbell (\cite{C87}), who we
will refer to now as W\&C, the toroidal component of the magnetic field is
written as being proportional to the relative angular velocity between the disc
and the star multiplied by the vertical field, which is taken to be a pure dipole
(see Eq.~(\ref{eq:bp_an})). For models where $\Omega_{\rm disc}$ is Keplerian
and the inner edge of the disc is inward of the corotation point, $B_\phi$ has a
positive global maximum at the inner edge of the disc, is zero at the corotation
point, reaches a global negative minimum somewhere outward of this and then
tends towards zero at very large $r$. If $\Omega$ deviates from Keplerian in the
inner part of the disc (reaching a maximum and then decreasing again as one
moves inwards), then the location of the maximum of $B_\phi$ is not in general
at the inner edge, but depends on the precise profile of $\Omega$ in this inner
region.

The above description is only partially in agreement with the results of our 
present two dimensional calculations. They share the feature of having a zero of
$B_\phi$ at the corotation point (or extremely close to it, see
Fig.~\ref{fig:multiple_theta_and_avg}). As regards the predicted global maximum
of $B_\phi$ in the inner part of the disc, all of our calculations show a
positive maximum close to where $\Omega$ has a maximum (compare
Figs.~\ref{fig:ref_cont_11.33}, \ref{fig:ref_eta_B3} and \ref{fig:eta_const_B3}
with Fig.~\ref{fig:omega}). However, when $\eta$ is not constant, magnetic field
lines accumulate and a second maximum can appear outward of the corotation
point, and this can also become a (positive) global maximum depending on the
value of $\eta_0$. Finally, regarding the minimum: as in the W\&C models, we
always find a global negative minimum before the outer edge of the disc;
however, when $\eta$ is not constant a second minimum can appear, thus producing
a structure with four extrema: two maxima and two minima (see
Fig.~\ref{fig:ref_cont_10.115}). We note here that even in the configurations
with only two extrema, as in W\&C, the profile of the toroidal component is
still different from that predicted by those models. In particular, the location
of the minimum and the magnitude of the field at both extrema can be very
different.

\begin{figure}[ht]
\centering
\includegraphics[width=0.45\textwidth]{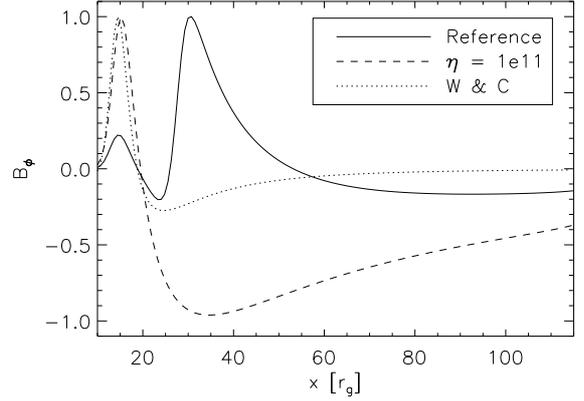}
 \caption{Comparison of three different profiles of the toroidal field: (1) the
solid curve is the shell average for the reference configuration, (2) the dashed
curve is the equivalent one for the configuration with constant $\eta$
($10^{11}$ cm$^2$ s$^{-1}$), (3) the dotted curve is given by the W\&C formula
with our $\Omega$ profile at $\theta = 88^\circ$. All of the curves have been
normalised so as to have their peak at $1$.}
 \label{fig:Bphi_cmp}
\end{figure}
\begin{figure}[ht]
 \centering
 \includegraphics[width=0.4\textwidth]{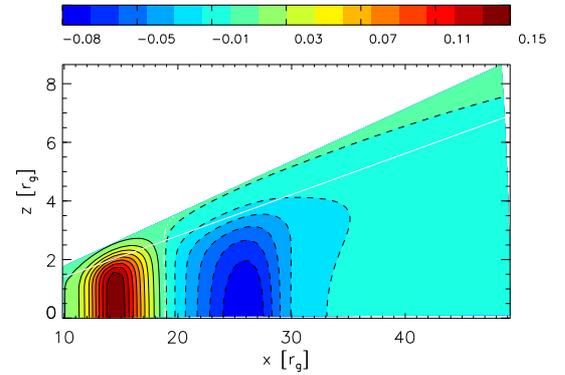}
 \caption{Contour plot of $\Delta\Omega \, B_\theta$ for the reference
configuration.}
 \label{fig:DWBt_ref}
\end{figure}

The main features of the toroidal field component, as given by our present 
numerical calculations and by the analytic models, can be seen in
Fig.~\ref{fig:Bphi_cmp}, where we show three profiles for $B_\phi$: the dotted 
curve is the W\&C profile for our model, as resulting from Eq. (\ref{eq:bp_an})
but where we have used our profile for the angular velocity near to the
equatorial plane (i.e. Eq. (\ref{eq:omega}) at $\theta = 88^\circ$), the solid
curve shows the shell average for our reference configuration, and the dashed
curve is for the configuration with constant $\eta$ ($10^{11}$ cm$^2$ s$^{-1}$).

The properties of having additional extrema of $B_\phi$ and of varying the 
location and magnitude of the two standard extrema, are not seen in the W\&C
models.

\subsection{Role of $B_p$ and $D_{\rm m}$}\label{sec:BPOL}

In Fig.~\ref{fig:DWBt_ref}, we show the quantity $\Delta\Omega \, B_\theta$,
where $\Delta\Omega = \Omega_{\rm disc} - \Omega_{\rm star}$ and $B_\theta$ is
the $\theta$-component of the poloidal field as obtained from our numerical
calculations for the reference configuration. This quantity has one global
maximum and one global minimum. The maximum is located very close to where
$B_\phi$ and $\Omega$ have their first maxima, and the minimum is at a radial
location close to that of the first minimum of $B_\phi$ (only a few $r_{\rm g}$
smaller - see Fig.~\ref{fig:ref_cont_11.33} and Table \ref{tab:min_max}). This
shows that the quantity $\Delta \Omega \, B_\theta$ is still relevant for
grasping the fundamental properties of the toroidal component of the field,
although care must be taken in choosing the profile of $B_\theta$. Some
differences between the predictions of the W\&C models and our numerical results
can, in fact, be explained in terms of the distortion of the $\theta$ component
of the magnetic field. However, in Fig.~\ref{fig:DWBt_ref} there is no evidence
for the additional maximum and minimum, and so this is not the whole story. We
need to consider the distortions of the field in more detail, and not focus only
on the quantity $\Delta\Omega \, B_\theta$.

\begin{figure}[ht]
 \centering
 \includegraphics[width=0.4\textwidth]{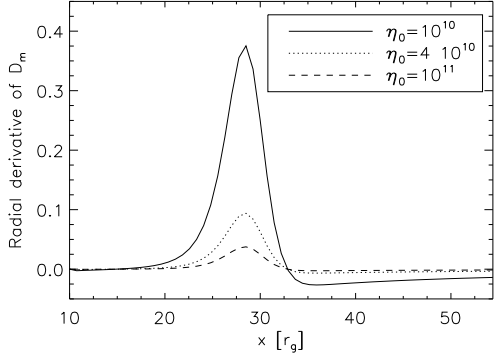}
 \caption{Radial derivative of the magnetic distortion function $D_{\rm m}$ 
near the equatorial plane (at $\theta = 89 ^ \circ$) for three configurations:
(1) the solid line is for the reference configuration with $\eta_0 = 10^{10}$
cm$^2$ s$^{-1}$, (2) the dotted line is for the configuration with $\eta_0 = 4
\, 10^{10}$ cm$^2$ s$^{-1}$, (3) the dashed line is for $\eta_0 = 10^{11}$
cm$^2$ s$^{-1}$.}
 \label{fig:der_Dm}
\end{figure}

In Paper I we have shown that the distortions of the poloidal component due to 
the plasma motion can be well described by a generalisation of the magnetic 
Reynolds number (which we have called the magnetic distortion function), 
defined as\footnote{Note that we have included $v_\theta$ here in the definition
of $D_{\rm m}$ (it was not present in Paper I, because we were showing there
results on the equatorial plane). We recall that $v_\theta$ is zero in the disc
and equal to $0.5\,v_r\,\tan\theta$ in the corona.}
 \begin{equation}
 D_{\rm m} = \frac{r_{\rm g}\,\sqrt{v_r^2+v_\theta^2}}{\eta} \, \rm{.}
\end{equation}
We studied $D_{\rm m}$ in detail in Paper I (see section 4 of that paper); here
we just recall that the magnitude of the peak in its radial derivative is a
measure of the degree of accumulation of poloidal field lines in its vicinity.
For the cases with constant $\eta$, the field amplification caused by this
distortion is negligible (the distortion is more homogeneous and the field lines
do not accumulate) and $D_{\rm m}$ is just proportional to $v_r$ in all of the
disc (so that there is no peak at all in the radial derivative). For the
reference case, instead, $\partial_r D_{\rm m}$ does have a peak and its radial
location ($r \sim 28.5 \, r_{\rm g}$) is near to that of the additional maximum
($r \sim 30 \, r_{\rm g}$). Increasing $\eta_0$ reduces the magnitude of the
peak and also the additional maximum of $B_\phi$ becomes less pronounced,
eventually disappearing for $\partial_r D_{\rm m} \leq 0.037$ (see
Fig.~\ref{fig:der_Dm}). We can therefore draw the conclusion that it is the
radial derivative of the magnetic distortion function which is responsible for
the additional maximum in the toroidal field profile.

\subsection{Our picture for $B_\phi$}\label{sec:BTOR}

In this subsection, we develop our alternative picture for the structure of the 
toroidal component of the magnetic field, following the same general approach 
as W\&C, but with a more detailed representation of the poloidal magnetic field 
and the velocity field, and a two-dimensional model. This is still a very 
simplified picture but we believe that it can represent a useful step forward, 
giving some additional insights. Our starting point is the $\phi$ component of 
the induction equation:
 \begin{equation}
 \partial_t B_\phi = [\nabla \times ( \vec{v} \times \vec{B})]_\phi -
[\nabla \times (\eta \nabla \times \vec{B} )]_\phi \, \rm{.}
\end{equation}
 The advective term is the one responsible for generating the field, while the 
diffusive term causes field losses. We then define the following scalar 
quantities:
 \begin{eqnarray}
 G \equiv \partial_t B_\phi|_+ &=& \left[ \nabla \times (\vec{v} \times
\vec{B}) \right]_\phi\\
 L \equiv \partial_t B_\phi|_- &=& \left[ \nabla \times (\eta \nabla \times
\vec{B}) \right]_\phi \, \rm{.}
\end{eqnarray}
 In a steady state $\partial_tB_\phi|_+= \partial_tB_\phi|_-$ so that $G = L$
and $\partial_tB_\phi=0$.

\begin{figure}[ht]
 \centering
 \includegraphics[width=0.45\textwidth]{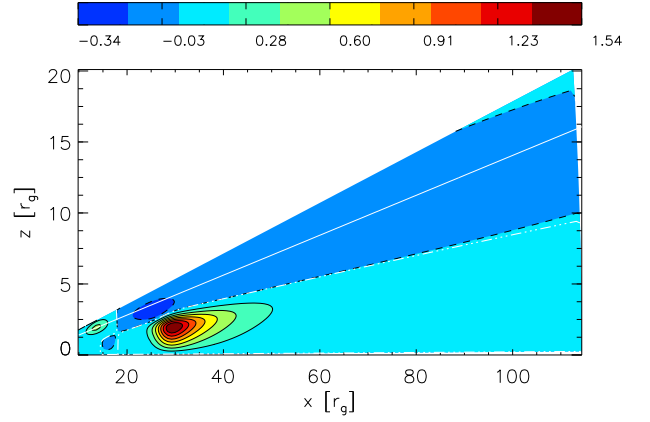}
 \caption{Contour plot of $a_0$ in Cartesian coordinates, in the region $r =
[10-115] \, r_g$. This should be compared with Fig. \ref{fig:ref_cont_10.115}.}
 \label{fig:a0_10.115}
\end{figure}

We focus first on the generation term and split it into two parts: one involving
the poloidal motion ($G_p$) and the other involving the toroidal motion
($G_\phi$):
 \begin{equation}
 \partial_tB_\phi|_+ = G_p + G_\phi \, \rm{.}
\end{equation}
 In spherical coordinates, these two terms are written as
\begin{eqnarray}
 G_p \equiv [\nabla \times (\vec{v_{\rm pol}} \times \vec{B})]_\phi &=& 
-\frac{1}{r}\left[ \partial_r(r\,v_r\,B_\phi) +
\partial_\theta(v_\theta\,B_\phi)  \right]\\
 G_\phi \equiv \left[\nabla \times (\vec{v_\phi} \times \vec{B})\right]_\phi
&=& \frac{1}{r} [ \partial_r ( r\,v_\phi\,B_r) + \partial_\theta (v_\phi \,
B_\theta) ]\, \rm{.}
\end{eqnarray}
 From these expressions we can see that, in general, the generation rate depends
on several quantities and not only on the vertical gradient of the angular
velocity (as in the W\&C models). In fact, all of the components of the velocity
field and magnetic field are present. However, usually in discs $v_\phi\gg
v_r,v_\theta$\footnote{In some circumstances one could have a strong wind from
the top surface of the corona and $v_r \ll v_\theta < v_\phi$.} and one can
expect the second term to dominate. An important difference between $G_p$ and
$G_\phi$ is that the former cannot generate any toroidal field from zero, but
can only modify $B_\phi$ once it has already been produced by some other means,
e.g. coming from advection by the azimuthal motion (i.e. from $G_\phi$).

We have calculated the ratio $G_\phi/G_p$ for our configurations and have found
that the toroidal term always dominates, even if the ratio is not as large as
would have been expected just by comparing the velocities (because one should
consider also the magnetic field). However even neglecting the contribution from
the poloidal advection, the generation rate for $B_\phi$ is still considerably
different from the one considered in the W\&C models. In fact, it contains also
the radial derivative of the term $r\,v_r\,B_r$ and the vertical derivative of
$B_\theta$.

\begin{figure}[ht]
 \centering 
 \includegraphics[width=0.4\textwidth]{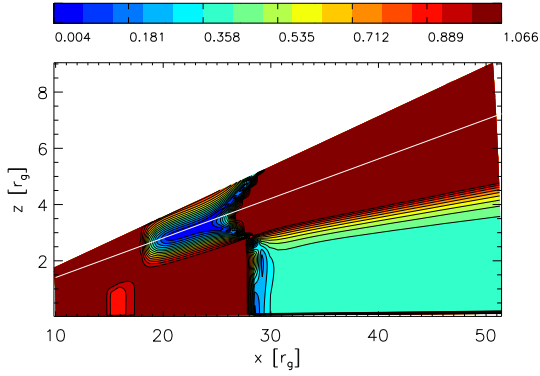} 
 \caption{Contour plot of $Q$ for the reference
configuration. All 
 locations with $Q > 1$ are marked in red so as to
highlight the regions 
 where $Q < 1$.}
\label{fig:RG}
\end{figure}

 So far, we have focused only on the generation term. In order to obtain the
profile of the toroidal field we have to equate it to the loss term, for which
we have
 \begin{eqnarray}
 L &\equiv& \left[  \nabla \times ( \eta \nabla \times \vec{B}  )  \right]_\phi
\\
&=& -\frac{1}{r} \left\{ \partial_r \left[ \eta \, \partial_r \left( r\,B_\phi
\right) \right] + \partial_\theta \left[
\frac{\eta}{r\,\sin\theta} \partial_\theta \left( B_\phi\,\sin\theta \right) 
\right] \right\} \, \rm{.}
\end{eqnarray}
 If we consider only typical values ($\tilde{\eta}$ and $\tilde{r}$), then we
can write
\begin{equation}
 L \sim \frac{\tilde{\eta}}{\tilde{r}^2}B_\phi \, \rm{.}
\end{equation}
 In a steady state, when $L=G\simeq G_\phi$, one has $G_\phi \simeq
(\tilde{\eta}/\tilde{r}^2) B_\phi$, and one gets
\begin{equation}
 \label{eq:bphi_tmp}
  B_\phi \simeq G_\phi\frac{\tilde{r}^2}{\tilde{\eta}} \, \rm{.}
\end{equation}
 In our configuration, $\eta$ has two main characteristic values: $\eta_0$ in 
the main disc region and $\eta_{\rm c} = 10^2 \, \eta_0$ in the corona and in 
the inner part of the disc. If in Eq. (\ref{eq:bphi_tmp}) we replace
$\tilde{\eta}$ with the actual profile of the magnetic diffusivity, and choose 
$\tilde{r} = r_g$, then we find that the variation of $B_\phi$ can be well 
approximated by that of the coefficient $a_0$ in Eq. (\ref{eq:B_short}):
 \begin{equation}
 \label{eq:my_bp}
 B_\phi \simeq \frac{r_g^2}{r\eta} \Big[ \partial_r ( r\,v_\phi\,B_r)
+ \partial_\theta (v_\phi \, B_\theta) \Big] = B_0 \, a_0
\end{equation}
 where $B_0$ is the reference unit of the magnetic field and we recall that in
Eq. (\ref{eq:B_short}) the fields are dimensionless (while here they are
dimensional).

 This formula for $B_\phi$ is rather well confirmed by our numerical
calculations. In Fig. \ref{fig:a0_10.115} we show the contour plot for $a_0$: 
this is very similar to that for the toroidal field (compare with
Fig.~\ref{fig:ref_cont_10.115}), the differences being due to the
approximations made in evaluating the loss term $L$, and to neglecting $G_p$ in
the generation term.

\subsection{Consistency check with the W\&C models}
 From Eq. (\ref{eq:my_bp}) we can see that the profile of the toroidal field is 
basically determined by four factors: (1) the radial derivative of $B_r$, (2)
the radial derivative of $r\,v_\phi$, (3) the vertical derivative of $B_\theta$
and (4) the vertical derivative of $v_\phi$.

In the W\&C models, the first three of these are neglected because $B_r$ is
taken to be zero everywhere and $B_\theta$ is supposed not to vary with
$\theta$. If one adds also the other assumptions used in their models (about 
the velocity profile and the disc thickness), one then finds that Eq. 
(\ref{eq:my_bp}) reduces to
 \begin{equation}
 B_\phi \sim \frac{\varpi}{h}\, (\Omega_K - \Omega_{\rm s}) \, B_\theta \,
\frac{r_g^2}{\eta}
 \end{equation}
 which is the same as Eq. (\ref{eq:bp_an}) with $\gamma_a = ({\varpi} / {h})$
and $\tau_d = ({r_g^2}/{\eta})$. We can therefore recover the expression
appearing in the earlier analytic models from our result. 

We now want to check a posteriori whether or not the simplifications made in the
W\&C models are still valid in our more general 2D model and, if so, in which
parts of the disc. One can do this by calculating the ratio ($Q$) between the
two terms on the right-hand-side of Eq.~(\ref{eq:my_bp}):
 \begin{equation}
\label{eq:RG}
 Q \equiv \frac{ |\partial_\theta (v_\phi \, B_\theta)| }  { |\partial_r
(r\,v_\phi\,B_r)| } \, \rm{.}
 \end{equation}
 Clearly, a necessary condition required for matching with the W\&C models is 
that $Q \gg 1$. In Fig.~\ref{fig:RG} we show a contour plot of this quantity for
the reference configuration. We show in red all of the regions where $Q > 1$,
while regions having the condition clearly being violated are colour-coded to
show by how much $Q$ is smaller than $1$.

There are large parts of the disc where this necessary condition is not met, 
with $Q$ even reaching values as small as $10^{-3}$. Moreover, even in regions
where $Q \gg 1$, the further necessary condition $\partial_\theta B_\theta \ll
\partial_\theta \Omega$ may not be met. In the main disc region the angular
velocity is almost constant with $\theta$ (the transition from Keplerian
rotation to corotation happens in the corona), and here we are exactly in an
opposite regime, i.e. $\partial_\theta B_\theta \gg \partial_\theta \Omega \sim
0$. Only in the corona and in the upper part of the disc are the vertical
gradients of $\Omega$ not negligible.

For a pure dipolar field, Eq.~(\ref{eq:RG}) becomes
\begin{equation}
\label{eq:RGdip}
 Q ^{\rm dip} = \left|\frac{1 + \frac{1}{2}
\cot\theta \frac{
\partial_\theta \Omega }{\Omega} } {-1 + r \frac{ \partial_r \Omega }{\Omega}}
\right|
\end{equation}
 which, for Keplerian motion, gives exactly $2/5$. Therefore the necessary
condition is never met for a pure dipole and Keplerian rotation, which is a good
description for the parts of our discs near to the mid plane. This should not
surprise us, since two key assumptions made in the W\&C models hold only in
different parts of our discs and not together: the field was supposed to have
both $B_r$ and $\partial_\theta B_\theta$ vanishing and decaying as for a dipole
(which we have only very close to the equatorial plane or for large $\eta$) and
the transition to corotation in the angular velocity was supposed to be very
sharp (which we have just beneath the disc surface, where $\Omega$ has to become
equal to $\Omega_{\rm s}$, far from the equatorial plane). We have calculated
$Q$ for a configuration with constant $\eta$ ($10^{12}$ cm$^2$ s$^{-1}$): in
this case the diffusivity is so strong that deviations away from the dipole are
quite small and so changes in $Q$ come almost entirely from the angular velocity
(according to Eq.~(\ref{eq:RGdip})). As expected, in the lower part of the disc
we find $Q \sim 0.4$.

Therefore having a larger value of $\eta$ does not necessarily imply that the 
necessary conditions hold in a larger region of the disc, it just implies that 
the field is closer to a dipole (which is what we are imposing at the
boundaries). In order to get $Q$ to go to infinity (which is what was assumed
in the W\&C models) not only does one need $B_r=0$ and $\partial_\theta B_\theta
= 0$ \footnote{These conditions hold exactly for a dipolar field only at the
equatorial plane; as one moves away from that, they are only partially
satisfied.} but also that the vertical gradient of the angular velocity, at the 
same location, has to be non-zero, or much larger than the vertical derivative 
of $B_\theta$.

\section{Conclusions}\label{sec:CON}
 In this paper we have considered a system consisting of a rotating neutron
star, having a dipole magnetic field aligned with the rotation axis and
surrounded by an accretion disc. The disc is truncated at the Alfven radius and
has a coronal layer above and below it. The region outside the corona is taken
to be vacuum and we impose dipolar boundary conditions on all of the boundaries.

Our aim was to improve on the analytic models developed by Wang (\cite{W87}) 
and Campbell (\cite{C87}) (W\&C). As in those models, we have made the kinematic
approximation and have looked for an axisymmetric stationary solution of the
induction equation, but we have gone beyond those models in solving for all of
the components of the magnetic field and not assuming the poloidal component to
be dipolar within the disc. We have also retained all of the components of the
velocity field rather than putting $v_r$ and $v_\theta$ to zero everywhere. We
have performed a fully two-dimensional calculation, without making any vertical
average or Taylor expansion in $h$ (the semi-height of the disc). Finally we
have neglected dynamo action but have included a turbulent magnetic diffusivity.

The analysis of the poloidal component of the magnetic field has been presented
in a previous paper (Naso \& Miller, \cite{papI}, Paper I); in the present paper
we have focused on the toroidal component. We have solved the $\phi$ component
of the induction equation numerically and have shown that the profile obtained
for $B_\phi$ can be very different from that in the earlier analytic models. 

In the W\&C models, the toroidal field strength was taken to be proportional to 
the relative angular velocity between the disc and the central object multiplied
by the vertical field, which was taken to be dipolar. However in Paper I we
found that, when calculated consistently, the poloidal field component was often
far from being dipolar. Therefore a first improvement with respect to the
earlier models was to use the poloidal field as obtained in our calculations,
i.e. a field dragged inwards by the plasma motion. This behaviour explains why
we then find different intensities for the toroidal field, and also a different
location for its global minimum.

When the turbulent magnetic diffusivity $\eta$ increases or the radial velocity
decreases one expects the field to be progressively less distorted by the plasma
motion. This is indeed what we have found both here and in Paper I. Our results
show that when the diffusivity $\eta_0$ is larger than about $10^{12}$ cm$^2$
s$^{-1}$, with the characteristic velocity $v_0$ being of the order of $10^5$ cm
s$^{-1}$, then the field is barely modified. Therefore whenever we expect the
magnetic field to deviate from the stellar dipole, $\eta$ should not be larger
than $10^{7} \, |v_{\rm cgs}|$ cm$^2$ s$^{-1}$, where $|v_{\rm cgs}|$ is the
characteristic magnitude of the radial velocity expressed in cm s$^{-1}$.

When the turbulent magnetic diffusivity is not constant throughout the disc and
corona, two additional extremal points may well appear: if the radial derivative
of the magnetic distortion function $D_{\rm m}$ is larger than a critical value
(about $0.04$ in the equatorial plane), there is an additional maximum and
minimum, and in some cases $B_\phi$ can even become positive again outward of
the corotation point, so that there are additional locations where $B_\phi=0$.
It is clear that under these circumstances the picture of which regions of the
disc tend to spin the star up or down has to be radically redrawn (this will be
the subject of a future investigation). However we should emphasise here that
there are still many uncertainties among experts about which profile of $\eta$
should be used and we have therefore made very simple choices here in line with
our step-by-step approach.

We have presented a new suggestion for the $B_\phi$ profile, which reduces to 
that of W\&C if one imposes $B_r = \partial_\theta B_\theta = 0$ and $\Omega =
\Omega_K$. In general there are large parts of the disc where the additional
terms included in our new picture for $B_\phi$ dominate over the one retained by
W\&C (see Fig.~\ref{fig:RG}). Our simplified expression (Eq.~\ref{eq:my_bp})
reproduces the numerical results quite well (compare
Figs.~\ref{fig:ref_cont_10.115} and \ref{fig:a0_10.115}), the differences being
due to approximations made in calculating the generation and loss terms for
$B_\phi$.

Summarising, in our calculations we have found that $B_\phi$ can have two maxima 
and two minima (see Fig.~\ref{fig:ref_cont_10.115}). The first maximum (positive
and inward of the corotation point) and the first minimum (negative and outward
of the corotation point) can be explained referring to the quantity $\Delta
\Omega \, B_\theta$, which has two extrema at the same locations as for $B_\phi$
(see Fig.~\ref{fig:DWBt_ref}). These extrema appear also in the W\&C models,
where the toroidal field is, in fact, taken to be proportional to $\Delta \Omega
\, B_z$. There is a fundamental difference however: in W\&C $B_z$ is a pure
dipole, whereas the $B_\theta$ which we consider here is that of a field being
dragged inward by the motion of the accreting material. 
 When $\eta$ is not constant, there is an additional maximum whose magnitude and
sign depend on the diffusivity in the disc, and whose radial location is always
outward of the first minimum, coinciding with that of the maximum in the radial
derivative of the magnetic distortion function $D_{\rm m}$. Outward of this
maximum, the field tends to come back to the profile that it would have had if
$\eta$ were constant, and this produces the last minimum (compare
Figs.~\ref{fig:ref_eta_avgs} and \ref{fig:eta_const_av}).

The main conclusion of this analysis is that, when the poloidal component of the
magnetic field is treated self-consistently in the calculations, the profile for
$B_\phi$ can be significantly different from that obtained by W\&C, and the
magnetic torque generated by it would then be different as well. Moreover, when
the turbulent magnetic diffusivity $\eta$ is not constant throughout the disc
and corona, some additional unexpected features can appear (such as a region of
positive $B_\phi$ outward of the corotation point). In the present work, we have
retained the very simple Keplerian rotation law in the main part of the disc.
Even within a purely hydrodynamical treatment, more complicated velocity fields
than this are expected (see Kluzniak and Kita, \cite{KK2000}; Jiao and Wu,
\cite{JW11}) and further changes are expected when back-reaction from the
magnetic field on the velocity field is included. The effects of this will be
another topic for investigation in subsequent stages of our step-by-step
approach.


 \appendix
\section{Testing of the code}\label{app:TEST}
 In this Appendix we discuss some of the tests that we have performed on the
numerical code used to solve Eq. (\ref{eq:B_short}). For a description of the
Gauss-Seidel relaxation procedure and of the discretization scheme see Appendix
A.1 of Paper I.

Before describing the tests, we should underline a difference in the boundary 
conditions with respect to the code used for the poloidal analysis. In Paper I 
we were not imposing the dipolar boundary conditions on the magnetic field 
directly but rather on the magnetic stream function $\mathcal{S}$. The poloidal 
magnetic field was then calculated by differentiating $\mathcal{S}$. Because of 
this, $B_r$ and $B_\theta$ were not precisely dipolar on the boundaries. For 
the calculations in the present paper, we have introduced a row of ghost points,
where we set the field to be exactly dipolar, i.e. the poloidal component is a
pure dipole and the toroidal component is zero.

We divide the tests into two groups. In the first group we chose the
coefficients of Eq.~(\ref{eq:B_short}) in such a way that it was possible to 
find an analytic solution, while in the second group we used the values given 
by Eqs.~(\ref{eq:att})-(\ref{eq:a0}). Here is a schematic description of these 
tests:
 \begin{enumerate}
 \item Tests with analytic solutions\\
 In addition to fixing the coefficients, one also has to choose the boundary 
conditions. We considered three sub-cases:
\vspace{0.2cm}
 \begin{enumerate}
  \item[1.1] All coefficients constant and set to $1.0$. There is 
then the following analytic solution:
  \begin{equation}
    B_\phi=\exp(-x/10^2 -
\theta/2) \, \cos\left(\sqrt{2.9604} \, \theta/2\right)-1 \, \rm{.}
  \end{equation}
  \item[1.2] All coefficients set to $0.0$ except for
  \begin{eqnarray}
    && a_{tt} = 1/x^2\\
    && a_x = b/x \, \rm{.}
  \end{eqnarray}
  The general solution is then:
  \begin{equation}
     B_\phi = h \, x^{k_1} \exp \left( i \sqrt{k_2} \, \theta \right)
  \end{equation}
  where $k_1$ is a function of $b$ and $k_2$. We chose $h=10$, 
$k_2=36^2$ and $b=2-k_2$, so as to include a complete period of the 
angular part of the solution within our angular domain. The solution is 
then:
  \begin{equation}
    B_\phi = \frac{10}{x} \, \cos \left( 36 \, \theta \right) \, \rm{.}
  \end{equation}
  \item[1.3] The same choice of coefficients as in test 1.2 but with 
different boundary conditions: $h=10$, $k_2=0$ and $b=70$. The analytic 
solution is then:
  \begin{equation}
    B_\phi = 10\, \rm{.}
  \end{equation}
 \end{enumerate}
\vspace{0.3cm}
 \item Testing the model setup\\
 In this group of tests we used a setup which was very similar to that used for
our actual physical analysis but varied some numerical parameters so as to test
the code. We performed four tests aimed at:
\vspace{0.3cm}
 \begin{enumerate}
  \item[2.1] checking convergence by changing the number of grid-points;
  \item[2.2] studying the dependence of the solution on the 
location of the radial outer boundary;
  \item[2.3] studying dependence on the initial estimate for the 
solution;
  \item[2.4] optimising the iteration step size by simple benchmarking.
 \end{enumerate}
\end{enumerate}

\subsection{Tests with analytic solutions}
 For all of the tests in this category, we considered the code stability and
convergence. We used grids with different numbers of points in both directions
and compared the analytic errors and the solutions.

In all cases, the stability of the code was related to the size of the iteration
step, the code being stable for values smaller than a certain threshold.

To check convergence, we considered how the maximum of the analytic errors 
changed with varying the total number of iterations. We considered both 
absolute and relative errors and analysed them in the region of interest (i.e. 
for $r<r_{\rm lc}$) by calculating their maximum and looking at their 2D
profile. As the iteration procedure continues, the errors decrease and at a 
certain point they saturate, so that making more iterations no longer leads to 
smaller errors. In addition to the errors, we have also considered the evolution
of the root mean square (rms) of the solution.

The size of the errors at saturation depends on the grid resolution, being
smaller for grids with more points (the size of the domain is fixed, so that 
increasing the number of points means increasing the resolution). Moreover the 
improvement obtained when increasing the number of grid-points becomes
progressively smaller, as it should do in a convergent regime. We calculated an 
effective order of convergence $p_{\rm eff}$ by considering the maximum value 
of the relative error at the final iteration, and how it changed with the grid 
size. Doing this we obtained $p_{\rm eff} \sim 1.5$.

All of the tests gave satisfactory results, confirming stability and giving 
convergence within $\sim 10^4$ iterations for tests 1.1 and 1.2, and within 
$\sim 10^7$ iterations for test 1.3. The maximum saturation error was $\sim
10^{-9}\%$ for test 1.3 and $\sim10^{-3}\%$ for test 1.1. As regards test 1.2,
since the solution has some zeros in the considered domain, we could not
estimate the error by considering the relative error (since it diverges at the 
zeros). We instead considered the maximum of the absolute error and compared it 
with the rms of the solution, finding that it was less than $1\%$.

\subsection{Tests with the model setup}\label{app:test_mod}
 For these tests we had no analytic solutions, and so could not calculate 
analytic errors but only residuals. When considering stability, we looked at 
the residuals, while for testing convergence we considered the change in the 
$\theta$-average of the solution (or in its rms) when changing the number of 
grid-points.

We recall that we cannot change the number of grid-points freely. In fact, 
one of the coefficients ($a_0$) is not calculated from an analytic function but
comes instead from the numerical results of Paper I for the poloidal field. This
coefficient is therefore directly defined only on the grid used in that
calculation, which had $1001 \times 21$ points. We consider this grid as our
reference one. When using a grid with fewer points, we have to choose them as a
subset of our reference grid, while when using a larger number of points, we
have to interpolate. Similarly when changing the size of the domain (by reducing
$r_{\rm out}$), we must also decrease the number of grid points $N_i$
accordingly, as explained at the end of Sect.~\ref{sec:sol_met} (see also
Eq.~(\ref{eq:Ni})).

A very important result of test 2.1 is that if the transition in $\eta$ between 
the disc and the corona is not well-enough resolved, then we find convergence 
to a different solution. More precisely, using an angular width for the 
transition of $5 \times 10^{-2}$ radians we need to have at least $20$ grid 
points overall in the $\theta$ direction in order to converge to the correct 
solution, i.e. to the same solution as for grids with larger values of $N_j$ 
(we tested with $N_j=39$ and $N_j=77$). This gives a minimum number of angular 
zones required for convergence of about $5$ within the transition region.

Test 2.2 tells us that the solution is not very dependent on the location of 
the radial outer boundary, contrary to the situation in Paper I for the poloidal
field. The $\theta$-averaged solution obtained using $r_{\rm out} = 290$ differs
from that obtained with $r_{\rm out}=750$ by less than $0.02\%$ in all of the
region of interest. If we consider the solution rms, the difference is even
smaller, being less than $5 \times 10^{-4}\%$.

For test 2.3, we ran calculations with quite different, and even unphysical, 
initial estimates for $B_\phi$, in order to see whether the solution still 
converged correctly. We used (1) a constant value $B_\phi = 1$; two decaying 
profiles: (2) $B_\phi = (10/x)^{3}$ and (3) $B_\phi = (10/x)$; (4) a growing 
profile $B_\phi = \sqrt{x/10}$ and (5) a Gaussian profile in both directions 
(centred at $x = 100\,r_g$, $\theta = 85^\circ$ with widths $15\,r_g$ and 
$2^\circ$). We obtained the same final solution for all of the cases; profiles 
(3) and (5) converged after $\sim 2 \times 10^6$ iterations, while all of the 
others converged after $\sim 4 \times 10^6$ iterations.

Finally in test 2.4 we changed the iteration step size $\Delta t$, looking 
for the largest possible value still giving stability. As in Paper I, we 
found that the maximum value depends more sensitively on $N_j$ than on 
$N_i$. For $N_j=81$ we found $\Delta t_{\rm max} = 4.7476 \times 10^{-4}$.

\begin{acknowledgements}
 This work has been partially supported by CompStar, a Research Networking 
Programme of the European Science Foundation, and by the National Natural
Science Foundation of China (40890161, 2011CB811403, 11025315, 10873020 and
10921303). L.N. is currently supported by a Chinese Academy of Sciences fellowship for young international scientists (grant No. 2010Y2JB12) and would
also like to thank the Department of Physics (Astrophysics) of the University of
Oxford for support granted during the development of this work, CAMK (Centrum
Astronomiczne im. M. Kopernika) in Warsaw, which provided partial support
through Polish Ministry of Science grant N N203 381436, and SISSA (Trieste)
whose high performance computing facilities have been used for running the
numerical calculations.
\end{acknowledgements}

\end{document}